# Panacea or Placebo? Exploring Causal Effects of Nonlocal Vehicle Driving Restriction Policies on Traffic Congestion Using Difference-in-differences Approach

Yuan Liang [a], Quan Yuan [a,b*], Daoge Wang [c], Yong Feng [a], Pengfei Xu [a], Jiangping Zhou [d]

[a] *Urban Mobility Institute, Tongji University, 201804 Shanghai, China*

[b] *College of Transportation Engineering, Tongji University, 201804 Shanghai, China*

[c] *School of Automotive and Traffic Engineering, Jiangsu University, 212013, Zhenjiang, China*

[d] *Department of Urban Planning and Design, Faculty of Architecture, The University of Hong Kong, Hong Kong SAR, China*

## Abstract

Car dependence has been threatening transportation sustainability as it contributes to congestion and associated externalities. In response, various transport policies that restrict the use of private vehicle have been implemented. However, empirical evaluations of such policies have been limited. To assess these policies' benefits and costs, it is imperative to accurately evaluate how such policies affect traffic conditions. In this study, we compile a refined spatio-temporal resolution data set of the floating-vehicle-based traffic performance index to examine the effects of a recent nonlocal vehicle driving restriction policy in Shanghai, one of most populous cities in the world. Specifically, we explore whether and how the policy impacted traffic speeds in the short term by employing a quasi-experimental difference-in-differences modeling approach. We find that: (1) In the first month, the policy led to an increase of the network-level traffic speed by 1.47% (0.352 km/h) during evening peak hours (17:00-19:00) but had no significant effects during morning peak hours (7:00-9:00). (2) The policy also helped improve the network-level traffic speed in some unrestricted hours (6:00, 12:00, 14:00, and 20:00) although the impact was marginal. (3) The short-term effects of the policy exhibited heterogeneity across traffic analysis zones. The lower the metro station density, the greater the effects were. We conclude that driving restrictions for non-local vehicles alone may not significantly reduce congestion, and their effects can differ both temporally and spatially. However, they can have potential side effects such as increased purchase and usage of new energy vehicles, owners of which can obtain a local license plate of Shanghai for free.

---

[*] Corresponding author. Address: 4800 Cao'an Road, Shanghai 201804, China. Email: quanyuan@tongji.edu.cn

*Keywords:* Nonlocal driving restriction policy, Traffic demand management, Traffic congestion, Difference-in-differences modeling

# 1. Introduction

Traffic congestion plagues many cities around the world. Cities have implemented various traffic demand management (TDM) policies to address this problem. For example, London, Singapore, and Stockholm have levied congestion charges (Lehe, 2019) whereas Milan, Paris, and Berlin have established low emission zones (Holman et al., 2015). In developing countries, driving restriction policies (DRPs) have been regarded as a low-cost and politically acceptable TDM measure (de Grange and Troncoso, 2011; Gallego et al., 2013). License-plate-number-based DRPs were first implemented in Latin American cities and then extended to cities elsewhere. For instance, Beijing, China first implemented a license-plate-number-based DRP during the 2008 Olympics (Li and Guo, 2016; Sun et al., 2014). Other Chinese cities such as Tianjin, Hangzhou, and Chengdu have also adopted similar DRPs. Unlike these cities, Shanghai, as the most populous megacity in China, has implemented a unique nonlocal-license-plate-based DRP since 2002, i.e., the nonlocal vehicle DRP (NLVDRP). According to this policy, nonlocal license plate vehicles are not allowed to drive on the roads within designated peak hours and popular areas.

Motivations underlying DRPs include reducing congestion and alleviating air pollution (Cantillo and Ortúzar, 2014; Sun et al., 2014). However, although DRPs have been widely investigated in terms of their effects on air quality (Bigazzi and Rouleau, 2017), the literature on how they affect traffic congestion remains limited and needs to be enriched (de Grange and Troncoso, 2011; Li and Guo, 2016; Li et al., 2022; Liu et al., 2018; Mohan et al., 2017; Sun et al., 2014; Yang et al., 2018). On the other hand, compared to license-plate-number-based DRPs, NLVDRPs are more stringent in a sense. License-plate-number-based DRPs reduce the number of driving days per week (for example, under one-day-per-week rules, vehicles remain free to drive on the other four unrestricted weekdays), while NLVDRPs permanently restrict the use of all targeted vehicles each weekday. In this way, the use of nonlocal vehicles under the NLVDRPs is much more severely impacted than that of the vehicles with a certain digit number under the License-plate-number-based DRPs. Considering the differences, the research question on how NLVDRPs would impact traffic conditions deserves more in-depth research.

A reliable *ex-post* evaluation of how NLVDRPs affect traffic conditions is thus needed. It can help policymakers and planners better make *ex-post policy* adjustments and follow-up decisions. However, there are many challenges in empirically evaluating the impacts. First, many cities do not have adequate financial resources to maintain a monitoring system that produces real-time empirical data ([Liu et al., 2018](#)). Traditional induction loops and microwave detectors can only monitor specific road segments' traffic flow and instantaneous speed but not network-level traffic conditions. Although many cities publish the transport performance index (TPI), the data is only available for the whole city or a few key parts of the city. The lack of traffic condition data with refined spatial and temporal resolutions becomes one of the biggest obstacles to accurately evaluating the impacts of DRPs. Second, the scope of DRPs is often the entire city. It is difficult for researchers to have control and treatment groups. Researchers can only compare traffic conditions before and after the implementation of DRPs. But traffic speeds/volumes may have periodic changes and be affected by many unobserved factors. Isolating the effects of DRPs from many confounding factors would require high-quality research design and settings ([Gibson and Carnovale, 2015](#)).

Shanghai has implemented a new round of NLVDRPs during weekday peak hours on the inner-ring area's surface roads (previously on elevated highways) since May 6$^{th}$, 2021. In this paper, we observe a quasi-experiment provided by the adoption of this NLVDRP to evaluate the short-term causal effects of the policy on traffic congestion using a high-spatiotemporal-resolution TPI data set. Specifically, we first build a panel data set as the measurement of traffic congestion levels based on a 10-minute-interval TPI of 68 traffic analysis zones (TAZs) before and after the implementation of NLVDRPs. We then take the TAZs within the NLVDRP implementation area (inner-ring area) as the treatment group and others as the control group. By employing the difference-in-differences (DID) modeling approach, we examine whether and how the NLVDRP impacted traffic congestion in the first month after its implementation. In addition, we conduct a series of robustness checks to validate our empirical findings and explore the heterogeneous policy impacts across TAZs.

This contribution of this paper can be summarized as follows. First, and most importantly, we offer empirical evidence on the short-term congestion-relief effects of a TDM policy that focuses on *urban cores* and restricts a subset of existing vehicles during peak hours. While such policy received less attention in the literature, it could become more popular in the future given its relatively low operation cost and narrow scope of impacts. Using Shanghai's NLVDRP as a case study, this paper has the potential to deliver insights to a broader audience of researchers and practitioners across the world where similar programs are applicable. For instance, our findings can be drawn upon and referenced when assessing how the implementation of Low Emission Zones (LEZs), a measure deterring the use of polluting vehicles in the presence of certain emission standards in the city center, would affect traffic congestion in the short run. After all, these two types of policies are similar in terms of spatiotemporal scale and impact groups. Second, we add new evidence into the literature regarding the intra-day, spillover, and spatially heterogenous effects of DRPs on traffic congestion by taking advantage of highly spatially disaggregated and high-frequency TPI data. Third, this paper, to the best of our knowledge, first applies a quasi-experimental DID modeling approach to identify the effects of DRPs on traffic congestion, providing *causal* instead of correlational estimates and achieving a higher level of validity. The causal inference framework we used can provide lessons for other studies aiming to appraise the impact of various transportation policies.

The remainder of the paper is organized as follows. Section 2 presents the literature review, followed by a description of Shanghai's NVLDRP in Section 3. Section 4 introduces the research design. Section 5 details the model results. Section 6 discusses the results. Section 7 concludes.

## 2. Literature Review

Drawing on the literature on DRPs, we can find that the majority of them focused on the policies' effects on mitigating tailpipe emissions, travel behavior, public transportation, and vehicle ownership. Only a small number of studies examined how DRPs would affect traffic congestion. Although the research topic of this study is traffic congestion, we believe a brief review of DRPs' other effects is also needed because it can help to demonstrate how the strand of literature on DRPs is developed and provide supportive information for this study. Therefore, we will first give a short summary of how DRPs affect non-congestion outcomes, such as air quality, and then focus on reviewing the literature on the effects of DRPs on traffic congestion.

Following Davis (2008), a large number of studies have explored the effects of DRPs in improving air quality (Gallego et al., 2013; Li et al., 2018; Sun et al., 2014; Viard and Fu, 2015; Ye, 2017; Zhong et al., 2017). Some studies found DRPs are ineffective (Davis, 2008; Sun et al., 2014; Ye, 2017) and others disclosed that DRPs could indeed mitigate emissions (Gallego et al., 2013; Viard and Fu, 2015; Zhong et al., 2017). It was also documented that as the purchase of additional vehicles and noncompliance behaviors emerge, the positive effects of DRPs would be eliminated or even go negative in the long term (Cantillo and Ortúzar, 2014; Davis, 2008; Gallego et al., 2013). In addition to air quality, researchers examined the impact of DRPs on travel behavior. Based on the 2010 Beijing Household Travel Survey (BHTS), Wang et al. (2014) found that 47.8% of the regulated drivers didn't follow Beijing's DRP and the policy had an insignificant impact on individual's decisions to drive, as compared to the DPR's impact on public transit. Using the same survey data, Gu et al. (2017) demonstrated Beijing's DPR had significant heterogeneous negative effects on auto trip frequency and thus VMT across subgroups of drivers. In contrast, Guerra and Millard-Ball (2017) found Mexico City's DPR had little-to-zero effects on overall vehicular travel. As for mode shift behavior, Liu et al. (2016) found that DRPs alone cannot effectively promote commuters to shift from personal cars to public transport if the policymakers fail to improve the service level of public transport. Zhang et al. (2019) found DRPs increased public transport ridership by 5-25% in six cities in China based on a panel data-based policy effect evaluation.

**Table 1** offers a review of empirical studies on the effects of DRPs on traffic congestion. The context includes Chile, India, and China. Three types of DRPs, namely catalytic-converter-based, license-plate-number-based one-day-per-week, and odd-even, were explored. Researchers leveraged various kinds of data to measure traffic congestion. Specifically, de Grange and Troncoso (2011) and Li and Guo (2016) used hourly detector-based traffic volume and speed data; Mohan et al. (2017) collected primary data from observational surveys; Liu et al. (2018) introduced disaggregate license recognition data into their study; Sun et al. (2014) and Yang et al. (2018) employed Beijing's city-level floating-vehicle-based TPI data; Li et al. (2022) leveraged segment-level traffic speed data released by a leading transportation network company. Technically, floating-vehicle-based TPI data provides higher traffic state recognition of the road networks than detector-based data (De Fabritiis et al., 2008). However, the public floating-vehicle-based TPI data is usually with a rough spatio-temporal resolution. For example, Beijing's public TPI data is daily updated for the whole city. No studies have yet applied TPI data with refined spatio-temporal resolutions to investigate the effects of DRPs on traffic congestion. Furthermore, since most DRPs target at the whole city area, traffic conditions lack DRPs-related variation at the spatial level after the deployment of DRPs. Researchers mainly utilized Ordinary Least Squares (OLS), Regression Discontinuity Design (RDD), or *t*-test to capture the variation of traffic speeds/volumes between days of restricted and unrestricted (de Grange and Troncoso, 2011; Li et al., 2022; Liu et al., 2018; Mohan et al., 2017). Besides, Sun et al. (2014) and Yang et al. (2018) used the daily variation in the number of cars being restricted due to the Chinese cultural avoidance of the number four as their identification strategy. Most of these studies indicate license-plate-number-based DRPs significantly reduced traffic volume and improved traffic speed. And the magnitude of these effects varied across different contexts. However, the effects of TDM measures analogous to NLVDRPs on traffic congestion are still not portrayed.

Table 1 Review of empirical studies on the effects of DRPs on traffic congestion.

| Reference | Context | Scheme of DRP | Measurement of traffic congestion | Spatial and temporal resolution | Modeling approach | Empirical findings |
|---|---|---|---|---|---|---|
| de Grange and Troncoso (2011) | Santiago, Chile | Catalytic-converter-based and license-plate-numbers-based | Traffic flow volume (the number of cars passing per hour) | 46 arcs in the Santiago road network; hourly | OLS | ● The permanent catalytic-converters-based DRP had no significant impact on traffic flow volume<br>● The additional license-plate-numbers-based DRP reduced traffic flow volume by 5.5% |
| Sun et al. (2014) | Beijing, China | License-plate-number-based one-day-per-week | Transportation performance index which is calculated using more than | The whole city (within 5th Ring Road); daily | OLS and robust OLS | ● More stringent DRPs had a positive effect on city-wide traffic speed |

| | | | 30000 floating vehicles | | | |
|---|---|---|---|---|---|---|
| Li and Guo (2016) | Beijing, China | License-plate-number-based odd-even | Traffic flow volume (pcu) and average speed (km/h) | 592 traffic detectors on the expressway network and some main arterials; hourly | Comparative analysis | ● During the DPR period, the traffic volume decreased by 20%-40% and the traffic speed increased by 10%-20% |
| Mohan et al. (2017) | Delhi, India | License-plate-number-based odd-even | Traffic volume and occupancy rates of vehicles derived from observational surveys; traffic speed derived from Google Maps Distance Matrix API | Four locations and 38/66 origin-destination pairs in Delhi; daily and hourly | Comparative analysis | ● Car flow rates decreased by less than 20% during the DRP, but the rates of motorized two-wheelers, buses, and autorickshaws increased<br>● Car occupancy rates didn't rise significantly |

| Yang et al. (2018) | Beijing, China | License-plate-number-based one-day-per-week | Transportation performance index which is calculated using more than 30,000 floating vehicles | The whole city (within 5th Ring Road); daily and hourly (only for 2013) | OLS | • The DPR significantly improved traffic speed during the restricted hours and did not worsen it during the unrestricted hours<br>• The effects of DRPs on evening traffic speed were stronger over years |
| --- | --- | --- | --- | --- | --- | --- |
| Liu et al. (2018) | Langfang, China | License-plate-number-based one-day-per-week and odd-even | Traffic volume, number of vehicles, and travel intensity derived from massive license plate recognition data | 106 detectors on 47 intersections within the restriction area; daily and hourly (aggregated by the authors from the raw data) | $t$-tests and RDD | • The shift of DRPs from one-day-per-week to odd-even reduced traffic volume by 8.74% in a normal weekday<br>• After the shift, traffic speed increased by |
| | | | | | | 23.30%, 12.85%, and 8.21% during morning peak hours, evening peak hours, and off-peak hours, respectively |
| Li et al. (2022) | Xian, China | License-plate-number-based one-day-per-week | Traffic speed provided by a leading transportation network company in China | 732 road segments within the restricted area; 10-minute | RDD | • The DRP increased traffic speed by 15%-20% during peak hours |

## 3. Background of NLVDRPs in Shanghai

Shanghai, a world-class megacity and the economic center of China, has experienced massive urbanization and rapid motorization in the past four decades. According to the 2019 Shanghai Comprehensive Traffic Operation Annual Report ([SCCTPI, 2020](#)), there were 5.4 million private vehicles in Shanghai in 2019. The rapid increase in vehicle ownership has brought about serious traffic congestion problems. The China Urban Transportation Report ([Baidu, 2021](#)) shows that Shanghai is the fifth most congested city in China, with peak hours congestion index of 1.932 (representing the ratio of actual travel time to free-flow travel time). To combat the rampant congestion, Shanghai has implemented a series of TDM policies ([Feng and Li, 2018](#)). In terms of vehicle purchases, Shanghai has established a unique local license plate auction system since the 1990s. Residents can participate in the monthly auction to obtain local license plates. The final transaction price was about 90,000 RMB (equivalent to about $14000) on average in 2020. Even so, the winning rate of the auction was only 5%. In response, a large proportion of residents adopted nonlocal license plates as alternatives, given they are free and easy to be obtained. This resulted in that nearly one-third of vehicles in Shanghai were with nonlocal license plates (about 1.7 million). Against this background, the government has implemented NLVDRPs since 2002 to curb their use. With the increased traffic congestion, the NLVDRP was extended to a larger geographic scale and a longer time period. As a result, vehicles with nonlocal license plates were gradually restricted from driving on the road within designated times and areas.

As shown in **Fig. 1**, Shanghai's NLVDRP targeted only at elevated highways in the city center before 2021. On May 6th, 2021, it was extended to surface roads (as shown in **Fig. 2**) for the first time. Specifically, during weekday morning and evening peak hours (7:00-9:00, 17:00-19:00), nonlocal license plate vehicles are restricted from driving on the surface roads of the inner ring area. Using this latest round of NLVDRPs as a case study, this study aims to investigate whether and how the policy contributes to congestion relief in the short run. (Hereafter, NLVDRP refers to the NLVDRP for the inner ring area's surface roads).

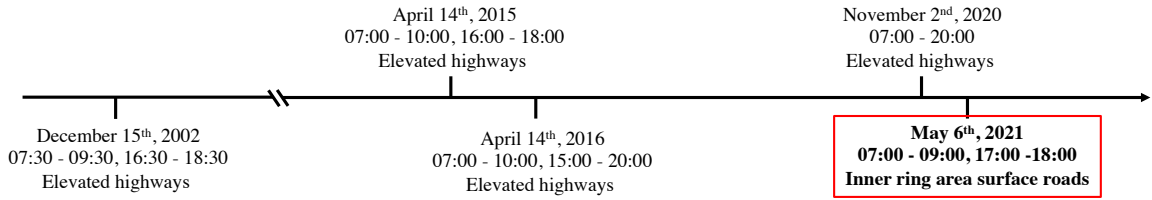

**Fig. 1** The timeline of Shanghai's NLVDRP.

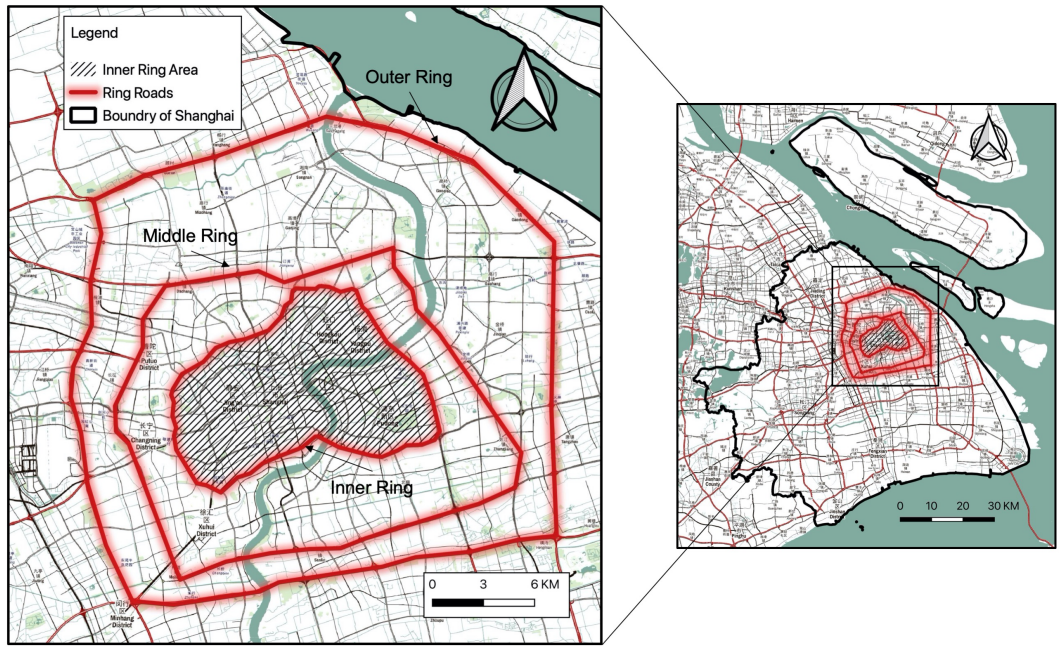

**Fig. 2** The inner ring area of Shanghai.

## 4. Research design

### *4.1 Data and study area*

To relieve the citywide traffic pressure during the 2010 Expo, Shanghai has set up a traffic congestion monitoring system with taxis as floating vehicles since then. After 11 years of development, the system has been constantly improved and updated. The real-time TPI data of 68 TAZs (as shown in **Fig. 3**) in the central city area is released to the public every 10 minutes through the website[2] from 6:00 to 23:00. The data is high-frequency and capable of capturing *network-*

---

[2] http://www.jtcx.sh.cn/trafficindex.html (in Chinese)

*level* traffic conditions, enabling us to measure traffic congestion in a more accurate, comprehensive, and objective way as compared to previous studies that relied extensively on segment-level traffic volume/speed data. Specifically, the TPI is calculated from the following equation:

$$TPI = \frac{\text{free-flow speed} - \text{actual speed}}{\text{free-flow speed}} \times 100 \qquad (1)$$

The actual speed is derived from the floating vehicles' real-time GPS data while the free-flow speed is set as the speed limit of road segments (Sun et al., 2016). The TPI of each TAZ is obtained by weighting the TPI of the road segments within the TAZ by length and number of lanes. The smaller the TPI, the closer the actual speed is to the free-flow speed, i.e., the higher the actual speed.

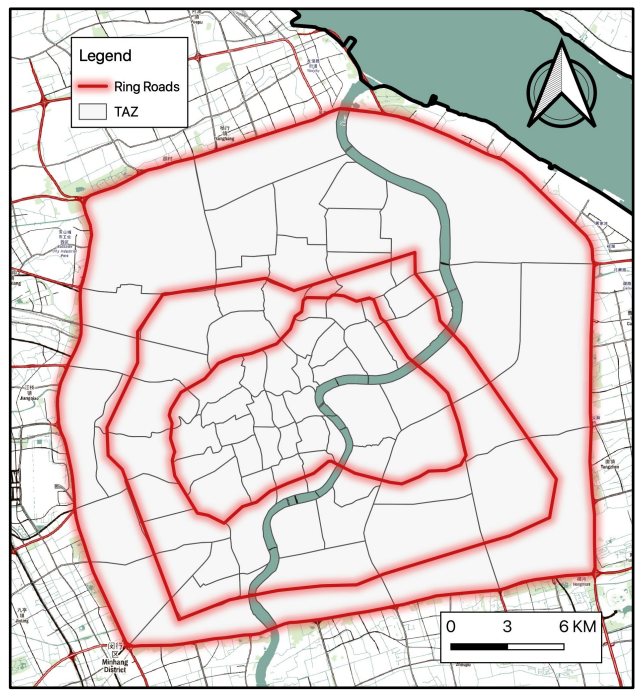

**Fig. 3** The TAZs in the central city area.

In this study, we collected the TPI data of all 68 TAZs for three weeks before and four weeks after May 6th, 2021 (the implementation of NLVDRPs) as the measurement of traffic congestion (as shown in **Fig. 4**).

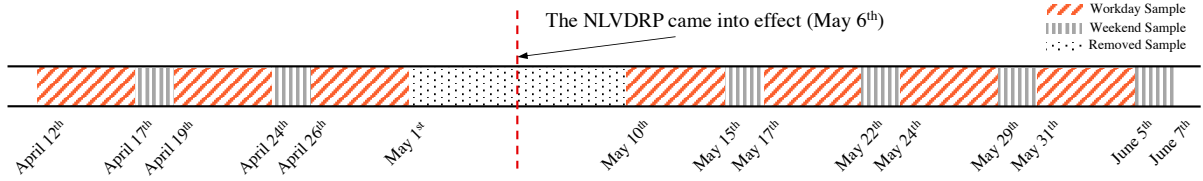

**Fig. 4** The timeframe of the sample.

Since May 1st to 5th is the Labor Day (an official holiday in China) and the travel pattern during that period may be different from usual, we remove this part of the TPI data. We also remove the TPI data from May 6th to 9th, as the travel pattern may not have immediately returned to normal after the long holiday. As a result, TPI data from April 12th to April 30th (pre-NLVDRPs) and May 10th to June 6th (post-NLVDRPs) is retained for the subsequent analysis. In addition, for the ease of understanding, we rescale the TPI data, which ranges from 0 to 100 as described in **Eq. 1**, by dividing it by 100. Such scale transformation allows the data to directly reflect the percentage deviation of the actual speed from the free-flow speed. We also review the data released by Shanghai Municipal Health Commission and find that there were no new confirmed COVID-19 cases or any COVID-19-control policy changes in Shanghai during the above period. Therefore, traffic congestion analyzed in this study is unlikely to be confounded by the pandemic. Furthermore, as shown in **Fig 5**, we divide 68 TAZs into four groups. TAZ_1 includes the TAZs within the inner ring area; TAZ_2 includes the TAZs which intersect with the inner ring area; TAZ_3 includes the TAZs adjacent to the inner ring area but not within it; TAZ_4 includes the TAZs that are neither within nor adjacent to the inner ring area. Given the NLVDRP only comes into effect in the inner ring area and nonlocal vehicles still can be freely driven outside the inner ring area, we assume that TAZ_4 is not affected by NLVDRPs and take it as the control group. TAZ_1 is the area of greatest interest and we take it as the focal treatment group. To test for the spillover effects of NLVDRPs, we also take TAZ_3 as the sub-treatment group. And we exclude TAZ_2 from our analysis since it crosses the boundary of the NLVDRP area.

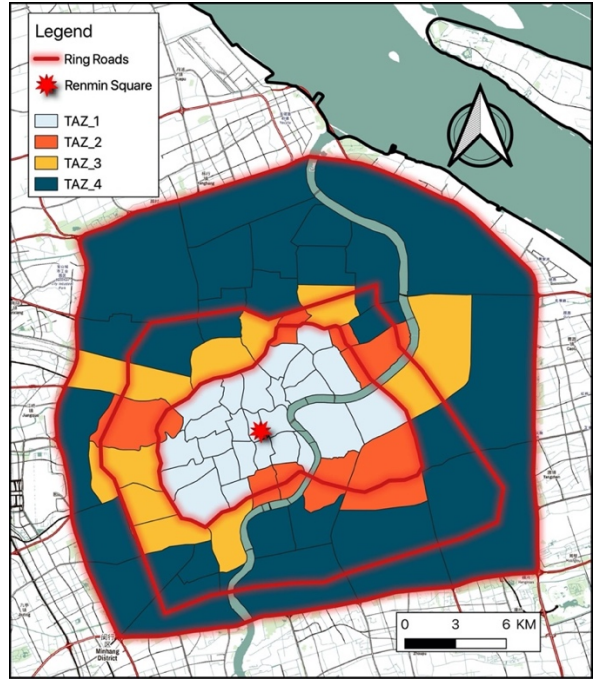

**Fig. 5** TAZ groups.

For each TAZ, we calculate the mean value of its daily morning and evening peak hour TPI separately as the outcome variable and compile a balanced panel data set. **Table 2** presents the descriptive statistics of the data set. **Fig. 6** shows the trends of the three TAZ groups' average TPI of weekday peak hours. It can be seen that before the implementation of NLVDRPs, the trends of the three groups are basically parallel.

**Table 2** Descriptive statistics of the TPI dateset.

|  | Mean (1) | SD (2) | Min (3) | Max (4) | Obs (5) |
|---|---|---|---|---|---|
| **Weekday** | | | | | |
| **Morning peak hours** | | | | | |
| TAZ_1 | 0.383 | 0.047 | 0.222 | 0.504 | 875 |
| TAZ_3 | 0.409 | 0.043 | 0.294 | 0.485 | 315 |
| TAZ_4 | 0.359 | 0.052 | 0.199 | 0.520 | 910 |
| **Evening peak hours** | | | | | |
| TAZ_1 | 0.450 | 0.054 | 0.300 | 0.592 | 875 |
| TAZ_3 | 0.419 | 0.052 | 0.286 | 0.555 | 315 |
| TAZ_4 | 0.349 | 0.055 | 0.147 | 0.498 | 910 |
| **Weekend** | | | | | |
| **Morning peak hours** | | | | | |
| TAZ_1 | 0.270 | 0.043 | 0.160 | 0.363 | 300 |
| TAZ_3 | 0.273 | 0.056 | 0.177 | 0.391 | 108 |
| TAZ_4 | 0.234 | 0.051 | 0.107 | 0.340 | 312 |
| **Evening peak hours** | | | | | |
| TAZ_1 | 0.403 | 0.066 | 0.187 | 0.559 | 300 |
| TAZ_3 | 0.359 | 0.053 | 0.250 | 0.494 | 108 |
| TAZ_4 | 0.300 | 0.058 | 0.139 | 0.415 | 312 |

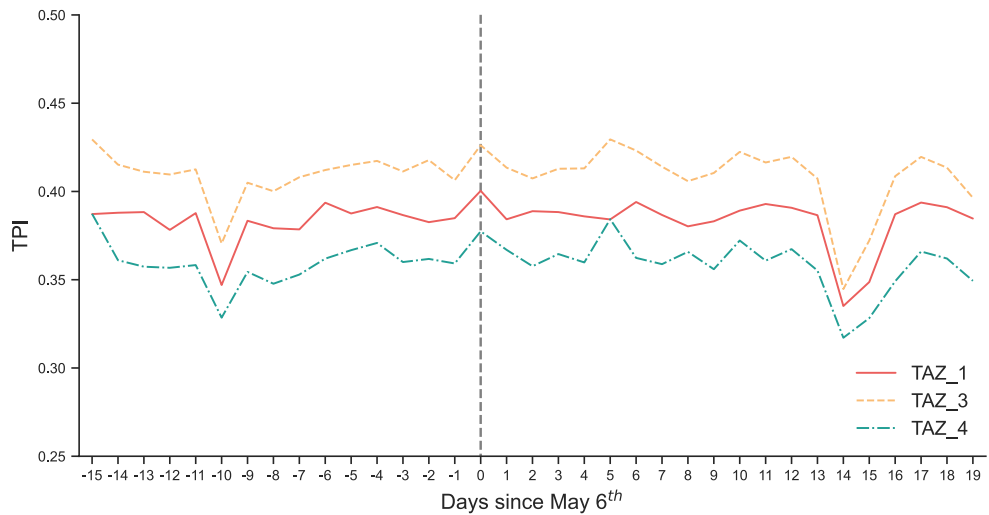

(a) Morning peak hours

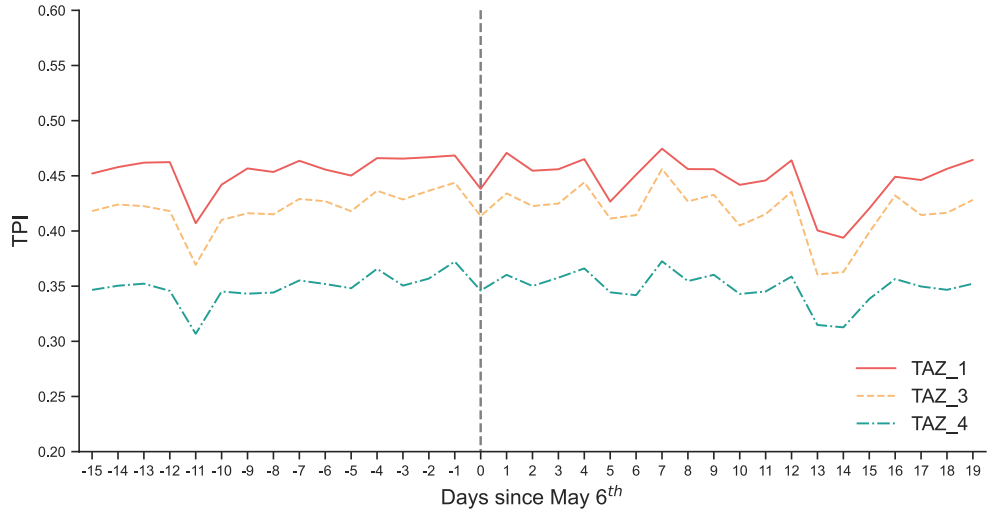

(b) Evening peak hours

**Fig. 6** The trends of the average TPI of weekday morning and evening peak hours.

*4.2 Empirical models and specifications*

Traffic congestion is subject to many spatiotemporally heterogeneous factors, including weather, special events, and so forth. Isolating the impacts of TDM policies on traffic congestion from many confounding factors is a challenging task. The quasi-experimental DID model is a commonly used approach to evaluate the causal effects of public policies on desirable output such as traffic congestion relief. Its central idea is presented in **Fig. 7**. The control group is assumed to be unaffected by the policy and thus can provide a counterfactual reference of what would have happened in the treatment group in the absence of the policy. We can obtain the average treatment effect of the policy by comparing the average changes in the outcome variables for the treatment and control groups after the policy implementation.

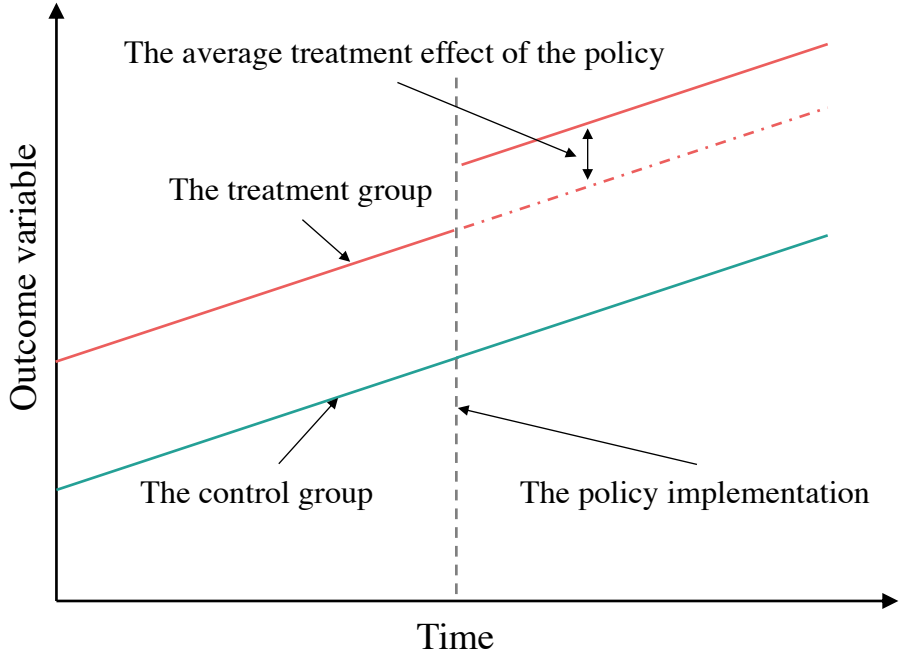

**Fig. 7** The graphical illustration of DID model

In this study, we utilize the following DID model:

$$TPI_{it} = \alpha_0 + \alpha_1 TAZ\_1_i \times Post_t + \alpha_2 TAZ\_3_i \times Post_t + \delta_i + \eta_t + \varepsilon_{it} \tag{2}$$

Where $TPI_{it}$ is the mean value of the TPI of the *i*-th TAZ on day *t*; $TAZ\_1_i$ is a dummy variable that takes the value of one if the *i*-th TAZ belongs to TAZ_1 and takes the value of zero otherwise; $TAZ\_3_i$ is the same. $Post_t$ is a dummy variable representing the implementation of NLVDRPs, which takes the value of one if the NLVDRP comes into effect on day *t* and takes the value of zero otherwise; $\delta_i$ is a set of dummy variables absorbing the time-invariant characteristics of each TAZ (TAZ fixed effects), including land use, population density, geographical characteristics, etc. $\eta_t$ is a set of dummy variables absorbing the characteristics of each calendar day (calendar day fixed effects), including weather, day of the week , special events, etc. That is, our model is a two-way fixed effect DID model. $\alpha_1$ and $\alpha_2$ are the coefficients of interest, representing the short-term causal effects of NLVDRPs on the TPI of TAZ_1 and TAZ_3, respectively. To account for potential inter-group error correlations, we cluster the standard errors at the TAZ level.

We also estimate the following DID model with a continuous treatment variable as the robustness check:

$$TPI_{it} = \alpha_0 + \alpha_1 DistanceToCityCenter_i \times Post_t + \delta_i + \eta_t + \varepsilon_{it} \qquad (3)$$

Where $DistanceToCityCenter_i$ is a continuous variable representing the distance from the $i$-th TAZ's geometric center to the city center (Renmin Square, as shown in **Fig. 5**); other terms are the same as in **Eq. 2**. $\alpha_1$ is the coefficient of interest. It measures how the NLVDRP affects the TPI gradient near the city center. We expect it to be positive, which means the policy reduces the TPI of TAZs near the city center (i.e., those in the inner ring area) and thus alleviates the traffic congestion.

Regarding the DID model, the most important assumption is that the outcome variables of treatment and control groups have common trends in the absence of the policy. Otherwise, the estimated effects of the policy may be biased. Although **Fig. 6** shows that the trends of the three TAZ groups' TPI are basically parallel, we formally test it by examining the following equation:

$$TPI_{it} = \alpha_0 + \sum_{k=\underline{k},\ k \neq -1}^{\bar{k}} \alpha_{1k} TAZ_{1i} \times Week_{kt} + \sum_{k=\underline{k},\ k \neq -1}^{\bar{k}} \alpha_{2k} TAZ_{3i} \times Week_{kt} + \delta_i + \eta_t + \varepsilon_{it} \qquad (4)$$

Where $Week_{kt}$ takes the value of one if day $t$ is in the $k$-th week before/after the implementation of NLVDRPs and take the value of zero otherwise; $[\underline{k},\bar{k}]$ is the range of time periods of the sample, where $\underline{k}$=-3 and $\bar{k}$=3; the week before the implementation of NLVDRPs is set as the reference ($k \neq -1$); other terms are the same as in **Eq. 2**. $\alpha_{1k}$ and $\alpha_{2k}$ are the coefficients of interest. We expect the coefficients of the time periods before the implementation of NLVDRPs are insignificant, which indicates that the difference in TPI between the treatment and control groups does not significantly change in the absence of the policy and thus the common trend assumption holds.

# 5. Results

## 5.1 Baseline results

Since the NLVDRP is only for weekday morning and evening peak hours, we estimate **Eq. 2** and **Eq. 3** with the corresponding sample as the baseline results and present them in **Table 3**. Besides, the estimated $\alpha_{1k}$ and $\alpha_{2k}$ of **Eq. 4** and their 95% confidence interval are shown in **Fig. 8** as the common trend test.

Table 3 Baseline results (Weekday model results)

|  | Morning peak hours (1) | Evening peak hours (2) | Morning peak hours (3) | Evening peak hours (4) |
|---|---|---|---|---|
| TAZ_1×Post | 0.001 | -0.008*** |  |  |
|  | (0.002) | (0.002) |  |  |
| TAZ_3×Post | -0.001 | -0.003 |  |  |
|  | (0.003) | (0.003) |  |  |
| DistanceToCityCenter×Post |  |  | 0.000 | 0.001*** |
|  |  |  | (0.000) | (0.000) |
| TAZ fixed effect | Yes | Yes | Yes | Yes |
| Calendar day fixed effect | Yes | Yes | Yes | Yes |
| No. of observations | 2,100 | 2,100 | 2,100 | 2,100 |
| Adj. R-squared | 0.929 | 0.955 | 0.929 | 0.955 |

*Significant at 5%, **Significant at 1%, ***Significant at 0.1%.

Robust standard errors clustered by TAZ in parentheses

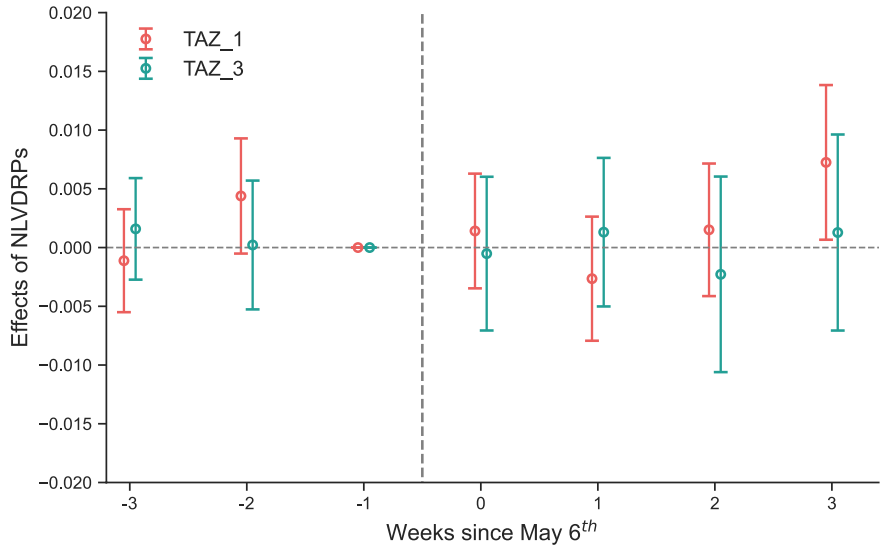

(a) Morning peak hours

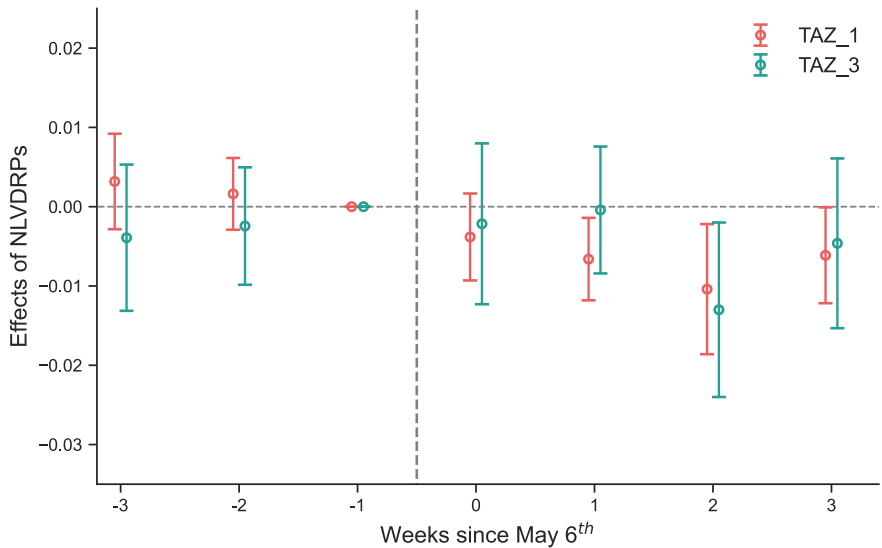

(b) Evening peak hours

**Fig. 8** Results of the common trend test

It can be seen that the effects of NLVDRPs on TAZ_1 and TAZ_3 are both statistically insignificant (95% confidence interval covers zero) before the implementation of NLVDRPs in **Fig. 8**, indicating the assumption of common trends is valid. Regarding the baseline results, the coefficients in column (1) of **Table 3** show that the effects of NLVDRPs are insignificant during weekday morning peak hours, both in TAZ_1 and TAZ_3. In other words, the NLVDRP has no significant impact on the traffic speeds of morning peak hours. In contrast, Column (2) shows that during evening peak hours, the NLVDRP has a significant negative impact on the TPI of TAZ_1 but not on TAZ_3. As mentioned above, the lower the TPI, the higher the traffic speeds are. Therefore, the NLVDRP significantly improves the traffic speeds of its target area (TAZ_1) during evening peak hours and has no spatial spillover effect. Columns (3) and (4) also confirm the above findings. The interaction term of *DistanceToCityCenter* and *Post* is only significant in the evening peak hour model and with a positive sign. This suggests that the NLVDRP has a negative impact on the TPI of the TAZs near the city center during evening peak hours but not morning peak hours, which is consistent with the results in columns (1) and (2). Furthermore, we could estimate the magnitude of the effects of NLVDRPs on traffic speed through the changes in TPI. Specifically, $-\Delta TPI \times free\text{-}flow\ speed^3$ represents the speed increased by the NLVDRP. Given the average free-flow speed of the inner ring area (TAZ_1) surface roads is about 44 km/h and $\Delta TPI$ is -0.008 during evening peak hours, the NLVDRP significantly improves the traffic speed by 0.352 km/h (=-(-0.008)*44km/h). Since the average traffic speed of the surface roads in the inner ring area during evening peak hours is about 23.966 km/h[4] before the implementation of NLVDRPs, the NLVDRP increased the traffic speed by 1.47% ((0.352 km/h)/(23.966km/h)).

*5.2 Robustness checks*

We first perform a placebo test as robustness checks. Specifically, we estimate **Eq. 2** and **Eq. 3** with the samples of weekend morning and evening peak hours. Since the NLVDRP does not restrict vehicles with nonlocal license plates on weekends, the estimated effects are supposed to be insignificant in the weekend models. **Table 4** shows the corresponding results. As expected, all the coefficients representing the effects of NLVDRPs on traffic congestion are insignificant, confirming the reliability of quasi-experimental DID modeling approaches, as well as the soundness of our baseline findings.

---

[3] $-\Delta TPI \times free\text{-}flow\ speed = -(TPI_{after} - TPI_{before}) \times free\text{-}flow\ speed = -(\frac{free\text{-}flow\ speed - actual\ speed_{after}}{free\text{-}flow\ speed} - \frac{free\text{-}flow\ speed - actual\ speed_{before}}{free\text{-}flow\ speed}) \times free\text{-}flow\ spee = -(\frac{actual\ speed_{before} - actual\ speed_{after}}{free\text{-}flow\ speed}) \times free\text{-}flow\ speed = actual\ speed_{after} - actual\ speed_{before}$

[4] The average TPI of the inner ring area TAZs before the implementation of NLVDRPs is 0.455328. Since $actual\ speed = free\text{-}flow\ speed \times (1 - TPI)$, the traffic speed is 23.966 km/h (44 km/h*(1-0.455328)). Also, according to Baidu's China Urban Transportation Report, the peak hour traffic speed of Shanghai (the whole city area) is 24.94 km/h in 2020, which is similar to our estimation.

**Table 4** Weekend model results

|  | Morning peak hours (1) | Evening peak hours (2) | Morning peak Hours (3) | Evening peak hours (4) |
|---|---|---|---|---|
| TAZ_1×Post | 0.001 | 0.002 | | |
|  | (0.004) | (0.003) | | |
| TAZ_3×Post | 0.005 | -0.000 | | |
|  | (0.005) | (0.003) | | |
| DistanceToCityCenter×Post | | | -0.000 | -0.000 |
|  | | | (0.000) | (0.000) |
| TAZ fixed effect | Yes | Yes | Yes | Yes |
| Calendar day fixed effect | Yes | Yes | Yes | Yes |
| No. of observations | 720 | 720 | 720 | 720 |
| Adj. R-squared | 0.900 | 0.943 | 0.900 | 0.943 |

*Significant at 5%, **Significant at 1%, ***Significant at 0.1%.

Robust standard errors clustered by TAZ in parentheses

In addition, since the NLVDRP is only for weekday peak hours, ideally, the NLVDRP should not affect traffic speeds during off-peak hours. However, DRPs may have inter-temporal substitution effects. That is, some drivers possibly change their departure time to avoid restriction hours and then reduce traffic speeds of other hours. To test this possibility and also for robustness checks, we run 17 sub-regressions of **Eq. 2** using weekday hourly samples (6:00-23:00). The corresponding coefficients are shown in **Fig. 9**.

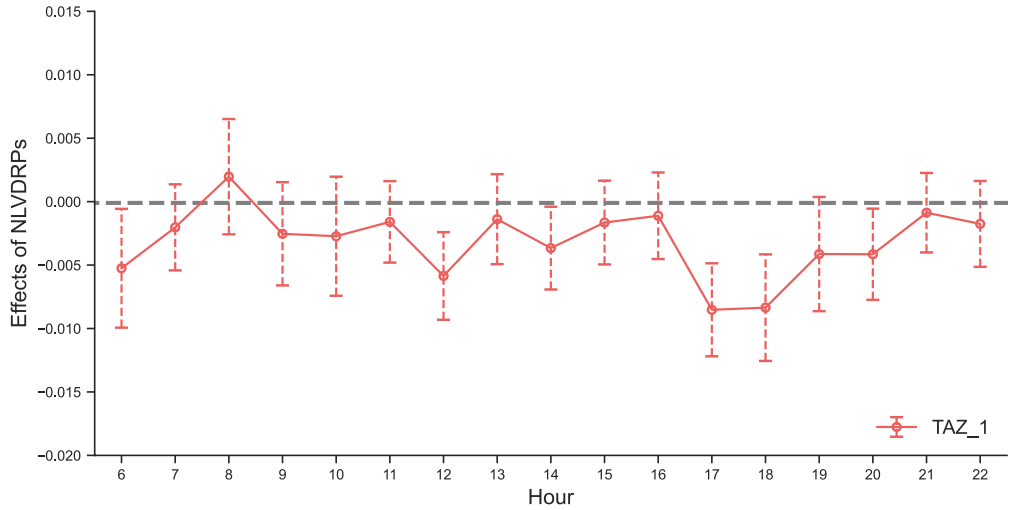

(a) TAZ_1

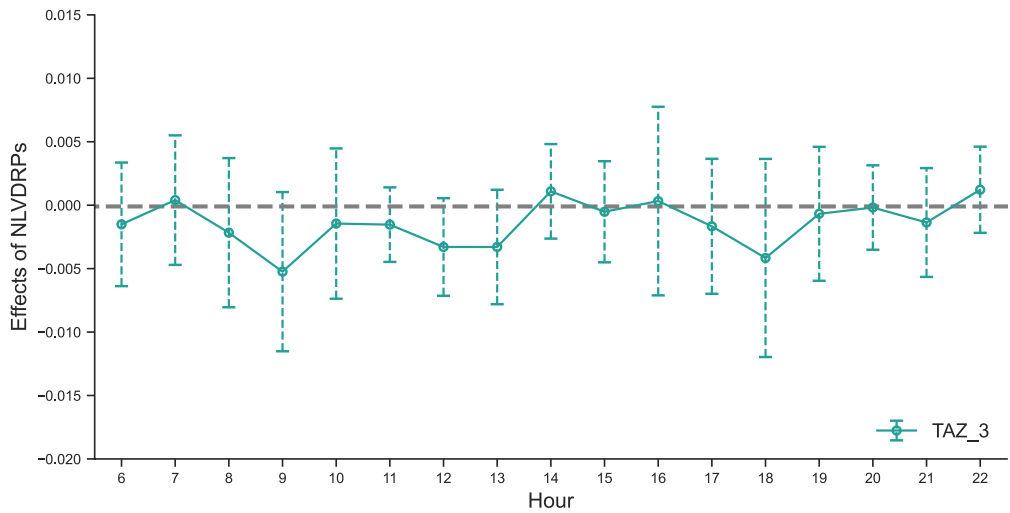

(b) TAZ_3

**Fig. 9** The temporal distribution of the effects of NLVDRPs

We can see that for TAZ_3, the effects of NLVDRPs are insignificant at any hour of the weekday. For TAZ_1, except for evening peak hours, traffic speeds are also slightly improved at 6:00, 12:00, 14:00, and 20:00, respectively. And the NLVDRP has no significant impact on traffic speeds during other hours. These findings further validate the robustness of our model results since the traffic speed of most off-peak hours is not affected by the NLVDRP. Furthermore, the results suggest that the NLVDRP had marginal inter-temporal positive effects on network-wide traffic speeds. Specifically, the NLVDRP improved the average traffic speeds of TAZs during some unrestricted hours.

In **Table 3**, TAZ_4 is set as the control group. We also conduct **Eq. 2** with alternative control groups as robustness checks. Specifically, we take the two hours adjacent to the weekday peak hours as the control group and only include the sample of TAZ_1. **Table 5** presents the model results. As we expected, the model results are consistent with the baseline results, indicating that the NLVDRP has significant negative effects on the TPI during evening peak hours but not morning peak hours.

**Table 5** Alternative control group results

|  | Control Group: 9:00-11:00 (1) | Control Group: 15:00-17:00 (2) |
|---|---|---|
| Morning_Peak_Hours×Post | -0.000 | |
|  | (0.001) | |
| Evening_Peak_Hours×Post |  | -0.007*** |
|  |  | (0.001) |
| TAZ fixed effect | Yes | Yes |
| Calendar day fixed effect | Yes | Yes |
| Hour fixed effect | Yes | Yes |
| No. of observations | 3500 | 3500 |
| Adj. R-squared | 0.819 | 0.863 |

*Significant at 5%, **Significant at 1%, ***Significant at 0.1%.
Robust standard errors clustered by TAZ in parentheses

*5.3 Heterogeneity analysis*

The above findings have shown that the NLVDRP has significant effects on traffic speeds of weekday evening peak hours. The highly spatially disaggregated nature of the TPI data allows us to further explore the spatial heterogeneity of such effects. Specifically, we estimate the following equation with the sample of weekday evening peak hours:

$$TPI_{it} = \alpha_0 + \alpha_1 TAZ\_1_i \times Post_t + \alpha_2 TAZ\_1_i \times Post_t \times X_i + \delta_i + \eta_t + \varepsilon_{it} \quad (5)$$

Where $X_i$ represents the built environment characteristics[5] of the $i$-th TAZ, including road network density, bus station density, metro station density, company density, tourist attraction density, and shopping and dining mall density. Other terms are the same as in **Eq. 2**. The model results are presented in Table 6, in which each column comes from a separate regression.

**Table 6** Heterogeneity analysis results

|  | (1) | (2) | (3) | (4) | (5) | (6) |
|---|---|---|---|---|---|---|
| TAZ_1×Post | -0.010* | -0.015** | -0.013*** | -0.006* | -0.007*** | -0.008*** |
|  | (0.004) | (0.005) | (0.003) | (0.003) | (0.002) |  |
| TAZ_1×Post×RoadDensity | 0.000 |  |  |  |  |  |
|  | (0.000) |  |  |  |  |  |
| TAZ_1×Post×BusStationDensity |  | 0.000 |  |  |  |  |
|  |  | (0.000) |  |  |  |  |
| TAZ_1×Post×MetroStationDensity |  |  | 0.004* |  |  |  |
|  |  |  | (0.001) |  |  |  |
| TAZ_1×Post×CompanyDensity |  |  |  | 0.000 |  |  |
|  |  |  |  | (0.000) |  |  |
| TAZ_1×Post×TouristAttractionDensity |  |  |  |  | -0.006 |  |
|  |  |  |  |  | (0.008) |  |
| TAZ_1×Post×ShoppingDinningMallDensity |  |  |  |  |  | 0.000 |
|  |  |  |  |  |  | (0.000) |
| TAZ fixed effect | Yes | Yes | Yes | Yes | Yes | Yes |
| Calendar day fixed effect | Yes | Yes | Yes | Yes | Yes | Yes |
| No. of observations | 2100 | 2100 | 2100 | 2100 | 2100 | 2100 |
| Adj. R-squared | 0.955 | 0.955 | 0.955 | 0.955 | 0.955 | 0.955 |

*Significant at 5%, **Significant at 1%, ***Significant at 0.1%.
Robust standard errors clustered by TAZ in parentheses

We observe that in Column (3), the triple interaction term is significantly positive, indicating the NLVDRP improved traffic speeds more in TAZs with less developed metro services. This suggests that the fraction of nonlocal vehicles might be greater in TAZs with low metro station density.

---

[5] The data is obtained from AMap Application Programming Interface (https://lbs.amap.com/) and the descriptive statistics of the built environment characteristics are provided in **Table A1**.

After the deployment of NLVDRPs, more vehicles were restricted from driving in these TAZs and thus traffic speeds were improved. Apart from that, we do not find other heterogeneous effects.

## 6. Discussions

In the empirical results, there are quite a few takeaways that are worth further discussing. In the light of baseline results, we found that the NLVDRP performed only in evening peak hours through increasing traffic speed by 1.47%. This suggests that this NLVDRP may not be as effective as license-plate-number-based DRPs in alleviating traffic congestion, given the evidence provided by previous studies (see **Table 1**). This may be because prior to the adoption of the policy, most nonlocal vehicles entering the inner ring area during morning peak hours were used for commuting. These nonlocal vehicle owners were more inelastic in travel decision making and were likely to take action in advance by switching to public transit or purchasing a local license plate ([Gibson and Carnovale, 2015](#)). If so, those nonlocal vehicles used for commuting on the surface roads of the inner ring area may disappear well before May 6$^{th}$, 2021. As a result, the immediate effects (effects compared to the most recent two to three weeks prior to the policy implementation, as indicated in the results above) of the NLVDRP on traffic speeds during morning peak hours turn out to be insignificant.

In fact, the implementation of NLVDRPs was announced as early as October 24$^{th}$, 2020. Thus, there were eight months for nonlocal vehicle owners to react in advance. On the other hand, during evening peak hours, there could be some nonlocal vehicles that travel for leisure and entertainment purposes after work and occasionally drive in the inner ring area. The NLVDRP effectively restricted these nonlocal vehicles from driving to and from the inner ring area and traffic speeds were improved due to the removal of these "more elastic" demand. In the meantime, given that these removed vehicles may only account for a small portion of all vehicles on the road, their impacts on traffic speeds were limited (1.47%).

Despite the above, we argue that the NLVDRP may promote urban sustainable transportation by boosting the adoption and usage new energy vehicles (NEVs). In order to facilitate the update of vehicle fleets towards electrification and sustainability, Shanghai offers local license plates to the buyers of NEVs for free. However, these plates for regular vehicles can only be obtained through public auctions at a very high price (about $14000),. Therefore, after the announcement of the NLVDRP, a large number of nonlocal vehicle owners chose to replace their nonlocal fuel vehicles with NEVs to avoid the driving restriction. At the same time, more and more residents are considering adopting NEVs as their first vehicles because of the high cost of obtaining a local license through public auction. **Fig 10** shows the historical Baidu index, which is similar to Google Trends, of NEVs in Shanghai. It represents the intensity of the search for NEVs on the largest search engine in mainland China. **Fig 11** further presents the historical sales of NEVs in Shanghai. It can be seen that the search index and sales of NEVs witnessed rapid growth after the announcement of the NLVDRP (October 2020), especially in the first three months. Furthermore, although we do not find any survey results regarding the intention to purchase NEVs in Shanghai, we still can draw some insights from existing studies conducted in other cities in China. For instance, Wang et al. (2017) and Li et al. (2020) indicated that the exemption from driving restrictions is a significant contributor to the adoption of NEVs, which is in line with our observations.

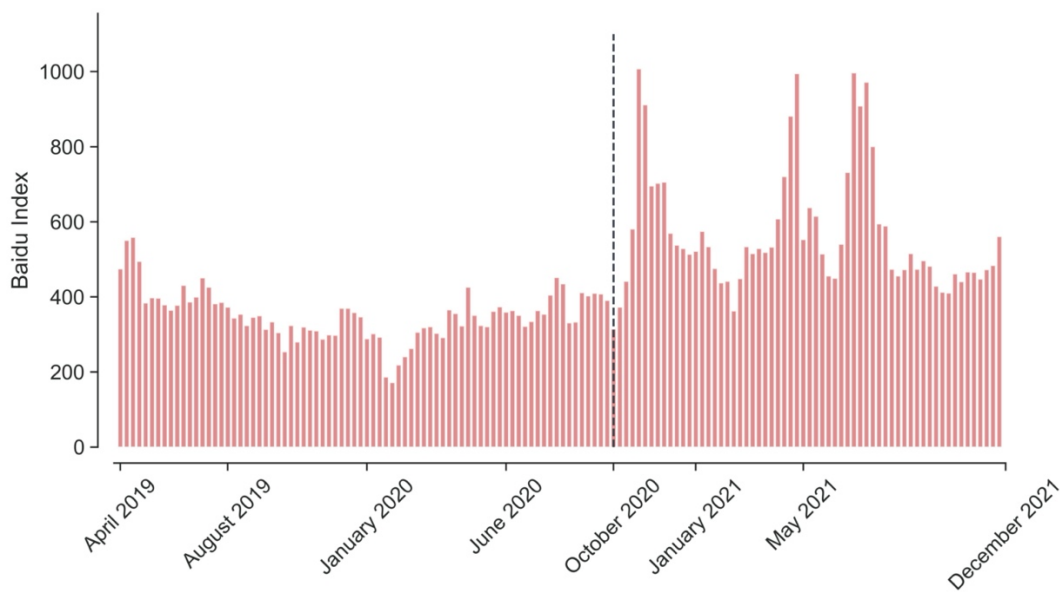

**Fig. 10** The baidu index of NEVs in Shanghai.

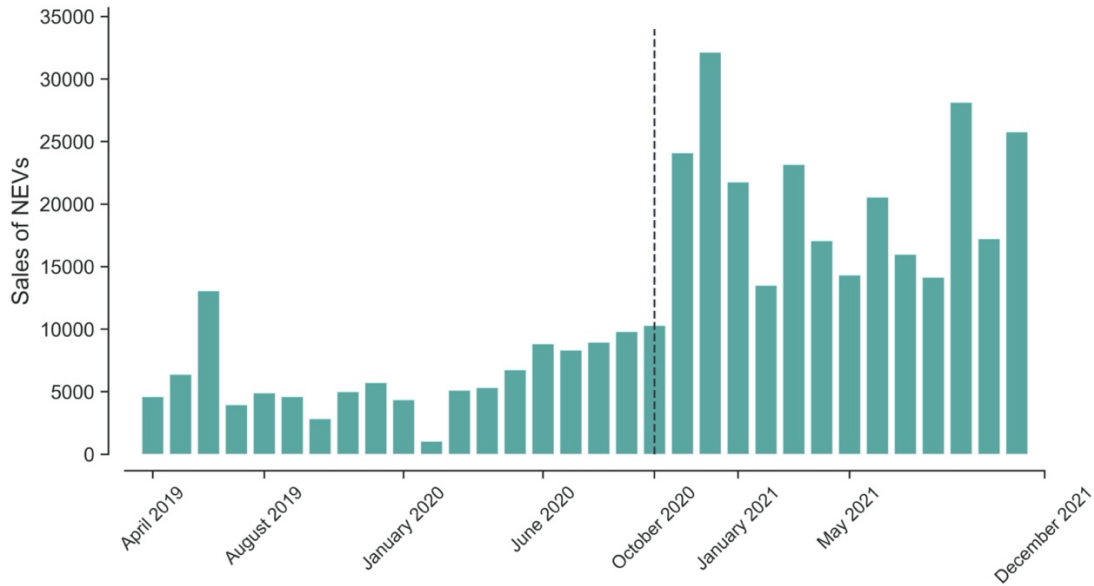

**Fig. 11** Sales of NEVs in Shanghai.

Last but not least, previous studies suggested that drivers who are restricted from driving by DRPs may adjust their travel schedules and routes (Mohan et al., 2017; Ye, 2017). For example, if the DRP is in place during morning and evening peak hours, restricted drivers might leave earlier or later to avoid the restriction. Also, they are likely to drive around the restricted area, i.e., make detours, during restricted hours. If so, traffic congestion in areas outside but adjacent to the inner ring, as well as that before or after peak hours, would become worse. But our empirical findings do not reveal this phenomenon as traffic speeds of TAZ_3 and during unrestricted hours do not significantly decrease after the implementation of the NLVDRP. One possible explanation is that the traffic speed was measured at the network-level in our study, while adaptive driving behaviors may only increase traffic volumes in some specific road segments. Therefore, we encourage future studies using multi-source traffic data, such as traffic count data and GPS trajectory data, to which we do not have access, to further decipher the effects of the NLVDRP on specific road types, such as arterials, sub-arterials, and so forth.

## 7. Conclusions

DRPs are usually considered effective and politically acceptable TDM measures to reduce traffic congestion in the short term (de Grange and Troncoso, 2011; Gallego et al., 2013). However, there is a lack of relevant empirical studies exploring their effects based on refined spatio-temporal resolution traffic condition data. Furthermore, while license-plate-number-based DRPs have received recurrent attention from scholars, NLVDRPs have not been well investigated yet. To fill these gaps, we use the 10-minutes-interval TAZ-level floating-vehicle-based TPI data to measure the average traffic speeds of the surface road networks of Shanghai and then evaluate the short-term causal effects of NLVDRPs on traffic congestion via a quasi-experiment design. According to the model results, we find that: (1) In the first month, the NLVDRP increased the network-level traffic speed by 1.47% (0.352 km/h) during evening peak hours but had no significant effects during morning peak hours. (2) In some unrestricted hours (6:00, 12:00, 14:00, and 20:00), the NLVDRP also increased the network-level traffic speed while the impact was marginal. (3) The short-term effects of NLVDRPs exhibit heterogeneity across TAZs. The lower the metro station density, the greater effects of NLVDRPs on traffic speeds were. Furthermore, the NLVDRP may promote the penetration of NEVs with free local license plates offered by the government.

Based on the empirical findings, we argue that NLVDRPs (or similar DRPs) are transitional TDM measures rather than final solutions to traffic congestion in megacities in developing countries. As revealed in this study, the NLVDRP in Shanghai has only generated marginal effects in reducing traffic congestion (only increasing traffic speed by 1.47% during evening peak hours on weekdays). Given the motorization in such megacities is still on the rise, further strengthening such DRPs may not be an effective and equitable option to address traffic congestion in the future. To solve this structural problem, policymakers need to adjust both the supply and demand sides. On the supply side, cities (especially megacities) in developing countries may consider further improving the *level-of-service* of existing roads through measures such as developing intelligent transportation systems and introducing more shared vehicles. The coverage and service quality of the public transit system should be simultaneously improved with the implementation of DRPs. On the demand side, the introduction of congestion charges or parking management policies might be helpful as vehicle ownership continues to grow. Furthermore, if NEVs can be exempt from driving restrictions whereas the purchase of the second fossil fuel car is restricted, the DRP can significantly encourage the ownership and usage of NEVs.

We have to acknowledge some limitations in this study as well. First, we only explore the short-term effects of NLVDRPs on traffic congestion. Nevertheless, the effects might be volatile after a long-term period. Future studies could focus on the long-term effects of this policy. Second, we only briefly discuss the effects of NLVDRPs on the promotion of NEVs, future research directions could explore this interesting topic with more convincing empirical evidence. Third, NLVDRPs could not only have an impact on traffic congestion, but also on air quality, parking demand, etc. It is worthwhile for us to investigate these issues in the future. Fourth, although it is beyond the scope of this paper, more studies in the future conducted in other cities may further shed light on our findings.

# Appendix

**Table A1** Descriptive statistics of the built environment characteristics of TAZs

| | | Distance to city center (in km) | Road density (in km per km$^2$) | Bus station density (in per km$^2$) | Metro station density (in per km$^2$) | Company density (in 100 per km$^2$) | Tourist attraction density (in 100 per km$^2$) | Shopping and dining mall density (in 100 per km$^2$) |
|---|---|---|---|---|---|---|---|---|
| TAZ_1 | Mean | 3.458 | 13.118 | 31.54 | 1.547 | 4.999 | 0.136 | 11.263 |
| | St. D | 1.684 | 3.93 | 7.843 | 0.798 | 2.586 | 0.127 | 6.331 |
| | Min | 0.406 | 8.195 | 19.059 | 0 | 1.685 | 0.011 | 2.707 |
| | Max | 6.946 | 23.2 | 54.361 | 3.124 | 13.843 | 0.59 | 32.874 |
| TAZ_3 and TAZ_4 | Mean | 9.858 | 7.866 | 17.909 | 0.395 | 1.314 | 0.017 | 3.028 |
| | St. D | 2.509 | 1.597 | 7.488 | 0.322 | 1.093 | 0.01 | 1.48 |
| | Min | 4.897 | 4.466 | 5.655 | 0 | 0.238 | 0.238 | 0.514 |
| | Max | 15.392 | 10.629 | 30.756 | 1.405 | 5.775 | 0.053 | 6.599 |

# Declarations

## Ethical Approval

Not applicable.

## Competing interests

The authors have no conflicts of interest or competing interests.

## Author's Contributions

Yuan Liang: conceptualization, data curation, visualization, methodology, software, writing - original draft. Quan Yuan: conceptualization, investigation, funding acquisition, writing- reviewing and editing. Daoge Wang: investigation, writing- reviewing and editing. Yong Feng: data curation. Pengfei Xu: data curation. Jiangping Zhou: writing- reviewing and editing.

## Funding

## Availability of data and materials

The data used in this study is available upon request.